\documentclass[conference]{IEEEtran}

\usepackage[noadjust]{cite}

\ifCLASSINFOpdf
   \usepackage[pdftex]{graphicx}
\else
   \usepackage[dvips]{graphicx}
\fi

\usepackage[cmex10]{amsmath}

\usepackage{algorithmic}

\usepackage{array}

\usepackage{mdwmath}
\usepackage{mdwtab}

\usepackage{eqparbox}

\usepackage[tight,footnotesize]{subfigure}

\usepackage{url}

\usepackage[font=footnotesize, justification=centering]{caption}

\begin{document}
\bstctlcite{IEEEexample:BSTcontrol}

\title{A Network Architecture for Distributed Event-Based \\Systems in an Ubiquitous Sensing Scenario}
\author{\IEEEauthorblockN{Cristina Mu\~{n}oz, Pierre Leone}
\IEEEauthorblockA{Department of Computer Science\\
University of Geneva\\
Carouge, Switzerland\\
Email: \{Cristina.Munoz, Pierre.Leone\}@unige.ch}}

\maketitle

\begin{abstract}
Ubiquitous sensing devices frequently disseminate their data between them. The use of a distributed event-based system that decouples publishers of subscribers arises as an ideal candidate to implement the dissemination process. In this paper, we present a network architecture which merges the network and overlay layers of typical structured event-based systems. Directional Random Walks (DRWs) are used for the construction of this merged layer. Our first results show that DRWs are suitable to balance the load using a few nodes in the network to construct the dissemination path. As future work, we propose to study the properties of this new layer and to work on the design of Bloom filters to manage broker nodes. 
\end{abstract}

\vspace{1em}
\begin{IEEEkeywords}
Distributed Event-Based systems; Directional Random Walks; Bloom filters; Wireless Sensor Networks.
\end{IEEEkeywords}

\section{Introduction}
\label{sec:introduction}

Ubiquitous or pervasive computing \cite{Cook:2012:RPC:2109687.2109848} uses many sources and destinations to gather and process data related to physical processes with the aim of making possible human-computer interaction. In the process of dissemination, some devices generate the data, while others are waiting for the sensing data. In this context, the use of a distributed event-based system \cite{Muhl:2006:DES:1162246} arises as an ideal candidate to implement the model of communication on the reception or transmission of events. 

\begin{figure}[!t]
\centering
\includegraphics[width=2.75in]{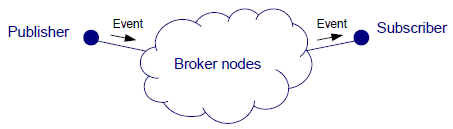}
\caption{Distributed notification service using a network of brokers.}
\vspace{-2.25em}
\label{PubSub}
\end{figure}

The main characteristic of an event-based system is that publishers and subscribers are decoupled. This means that they do not have any information about each other. The element in charge of matching notifications with subscriptions is called the event notification service. In distributed networks, the event notification service is implemented using a network of brokers nodes (see Figure~\ref{PubSub}). It is considered that a broker is any node in the network that has information about any single or set of subscriptions. The complexity of designing this type of systems usually lies on the way to elect the nodes which will act as brokers because of the decentralized nature of a distributed network.

In our research, we assume that a node can be a publisher, a subscriber, a broker or a combination of these three possibilities. We also assume that all the nodes in the network are able to participate in it without the requirement to adopt the specific role of publisher or subscriber. Nodes that are actively participating in the network but do not take any specific role will be considered as part of the overlay layer. Those nodes of the overlay layer that are able to redirect messages will be considered as brokers.

Event-based systems are classified as topic-based or content-based \cite{Muhl:2006:DES:1162246}. Topic-based systems take into account the subject of messages in order to match publications with subscriptions. Content-based systems use filters to specify the value of subscriptions' attributes to redirect notifications. A filter is a boolean function built taking into account the set of subscriptions. In our proposal, we plan to deal with a content-based system that uses Bloom filters at broker nodes in order to save memory resources and speed up routing decisions.

Sensor networks frequently use tiny devices with limited battery capabilities that make unsuitable the use of a Global Positioning System (GPS) to disseminate information according to the coordinates of nodes.  In addition to this, the use of virtual coordinates to substitute real coordinates requires the use of sinks or landmarks to structure the network. For these reasons, the use of coordinates in an unstructured sensing scenario is not recommended. We assume that we work in an unstructured scenario in which no routing protocol provides communication between the nodes of the network. 

The constraints of the network's infrastructure lead us to the design of a network architecture for distributed event-based systems that must use as less resources as possible (i.e., battery, memory, etc.). In this paper, we present a solution that avoids implying all the nodes of the network in the dissemination process by using a distributed notification service defined by Directional Random Walks (DRWs).

The rest of this paper is organized as follows: Section \ref{sec:state} analyzes the state of the art. Section \ref{sec:methodology} points out the approach to solve the problem specified in this section. Section \ref{sec:research} presents the research efforts already done for the approach specified in Section \ref{sec:methodology}. Finally, Section \ref{sec:conclusion} summarizes our proposal and suggests further work.

\vspace{0.2em}
\section{State of the art}
\label{sec:state}

\subsection{Distributed and Structured Event-based Systems}

\begin{figure}[!h]
\centering
\includegraphics[width=2.45in]{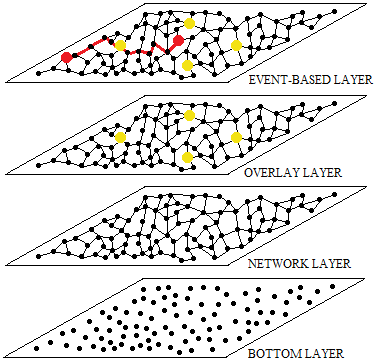}
\caption{Decomposition in layers of the typical design of a distributed and structured event-based system.}
\vspace{-1em}
\label{3layersdistributedusualsystems}
\end{figure}

Distributed and structured event-based systems use three layers on the top of a bottom layer (see Figure~\ref{3layersdistributedusualsystems}), which provides data link functionalities, to facilitate topology control:
\begin{enumerate}
\item The network layer is in charge of providing data forwarding between the different nodes involved in the network. A network protocol, such as the Multicast Ad-hoc On-demand Distance Vector (MAODV) \cite{Roy05securingmaodv:} is needed to provide point-to-point communication.
\item The medium layer is called the overlay layer. It is a virtual layer that builds the event notification service by providing a network of brokers that redirect notifications to the corresponding subscribers. 
\item Finally, on the top layer the event-based protocol is implemented.
\end{enumerate}

One strategy to construct the overlay layer is to use a tree. In TinyMQ \cite{shi2011tinymq}, which is designed specifically for wireless sensor networks, a multi-tree overlay layer is maintained. 

Another strategy is to clusterize the network and use cluster heads to manage messages as in Mires \cite{souto2006mires}, which is a middleware for sensor networks. The Gradient Landmark-Based Distributed Routing (GLIDER) \cite{Fang05glider:gradient} organizes the network using some defined landmarks to compute the Delaunay graph for network partition. Then, the Landmark-Based Information Brokerage scheme (LBIB) \cite{Fang06landmark-basedinformation} uses an overlay layer based in GLIDER to match publishers with subscribers. 

A typical solution is to build the overlay layer using Distributed Hash Tables (DHTs). In these systems, a key is mapped to a particular node with storage location properties. In some DHT architectures, rendez-vous nodes depend on the node ID as in Pastry \cite{Pastry}. In others, as the Content Addressable Network (CAN) \cite{CAN}, a region of the space is used to map a key. Some efforts have been made to apply this solution to sensor networks \cite{Fersi:2013:DHT:2429525.2429572}. When coordinates are available, sensor networks use Geographic Hash Tables (GHT) instead of a typical DHT. Currently, technology companies as Ericsson Research, are making an effort to develop applications that use GHTs in wireless sensor networks \cite{SENSORNETS14M3}.

\subsection{Distributed and Unstructured Event-based Systems}

The main characteristics of distributed and unstructured event-based systems is that they do not maintain an overlay layer. This fact makes easier to deal with network's changes. The distributed notification service may be built using flooding, gossiping or random walks.

Most of the algorithms proposed deal with the unstructured nature of wireless communications using flooding to build a tree. A typical solution is to use the On-Demand Multicast Routing Protocol (ODMRP) \cite{Lee:2002}, which is based in the forwarding group concept. Groups are constructed and maintained periodically when a multicast source has data to send. This task is done by broadcasting the entire network with membership information. An extension for ODMRP has been proposed \cite{Yoneki:2004} to adapt a content-based system by adding subscriptions to Bloom filters. Trees also may be configured to self-repair themselves in base to brokers' dynamicity \cite{Mottola:2008}. These solutions are reliable but increase the traffic of the network because they use flooding at some point.

Flooding may also be used to continuously exchange subscription information clusterizing the network \cite{Voulgaris06}. Then, notifications are sent to the appropriate cluster, improving the efficiency of the network. Other mechanisms can be used as the combination of a DHT and random walks \cite{Tian:2005}. Cluster heads manage the DHTs while random walks help to connect the different cluster heads of the network. The cluster concept in the network of brokers can be improved in a dynamic scenario by enriching the topology management with predictions based on location \cite{Abdennadher:2013:APM:2508222.2508234}. 

\subsubsection{Probabilistic approaches}
Probabilistic approaches are suitable to deal with dynamicity but they do not offer reliability. Some solutions propose that all the nodes in the network implement a broker that forwards messages to neighbors depending on the estimation of potential subscribers \cite{Haillot:2008}. Other solutions \cite{1437119}, propose to flood subscriptions in a small area and then use random walks to reach that areas. In Quasar \cite{Wong:2008}, subscriptions of a certain area are able to attract or reject notifications, that are propagated with a random walk, using an attenuated Bloom filter \cite{5751342}. A probabilistic solution that uses a random walk specifically designed to go deep into the network is CoQUOS (Continous Queries on Unstructured OverlayS) \cite{Ramaswamy:2011}. Continuous queries are launched to the network using random walks. Peers compute the overlap between their neighbor's lists and use this information to forward the random walk to avoid remaining in a cluster. Then, some peers register the query with a probability that depends on the number of hops.







\vspace{0.5em}
\section{Network architecture}
\label{sec:methodology}

Due to the unstructured nature of our network, we propose the development of a dissemination algorithm that merges the network and the overlay layers of a typical distributed and structured event-based system (see Figure~\ref{3layersdistributedusualsystems}). This means that no network protocol is needed. The main advantage of not using a network protocol is that there is no necessity of maintaining a network topology. This implies that most nodes of the network, which do not actively participate in the process of dissemination, do not have to keep any information about topology. The main consequence is that nodes not involved in the system are able to save energy and computing resources. 

\begin{figure}[!h]
\centering
\includegraphics[width=2.5in]{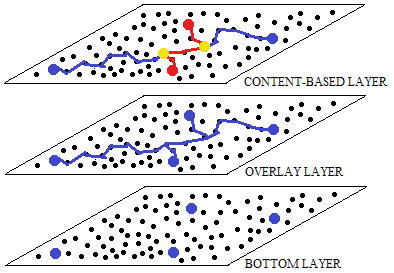}
\caption{Decomposition of the architecture of our design in layers.}
\vspace{-1.00em}
\label{3layers}
\end{figure}

Our design (see Figure~\ref{3layers}) uses two layers on the top of a bottom layer that provides data link functionalities:

\begin{enumerate}
\item The overlay layer is in charge of providing the distributed network of brokers and, at the same time, provides point-to-point communication between publishers and subscribers. The main objective of this strategy is to avoid the use of global information of the network which is costly to get and maintain.
\item As in Figure~\ref{3layersdistributedusualsystems}, the event-based protocol is implemented at the top layer.
\end{enumerate}

As Section~\ref{sec:introduction} mentions, we assume that a node can be a publisher, a subscriber, a broker or a combination of these three possibilities. Moreover, our design takes advantage of some nodes in the network which want to collaborate. Nodes that participate in the system are considered part of the overlay layer (blue path of Figure~\ref{3layers}). The overlay layer is formed by the intersection of different publishers and subscribers (blue nodes). Publishers and subscribers implement a DRW until intersecting other DRW. Broker nodes (yellow nodes) are the meeting point between two DRWs.

\begin{figure}[!h]
\centering
\includegraphics[width=2.5in]{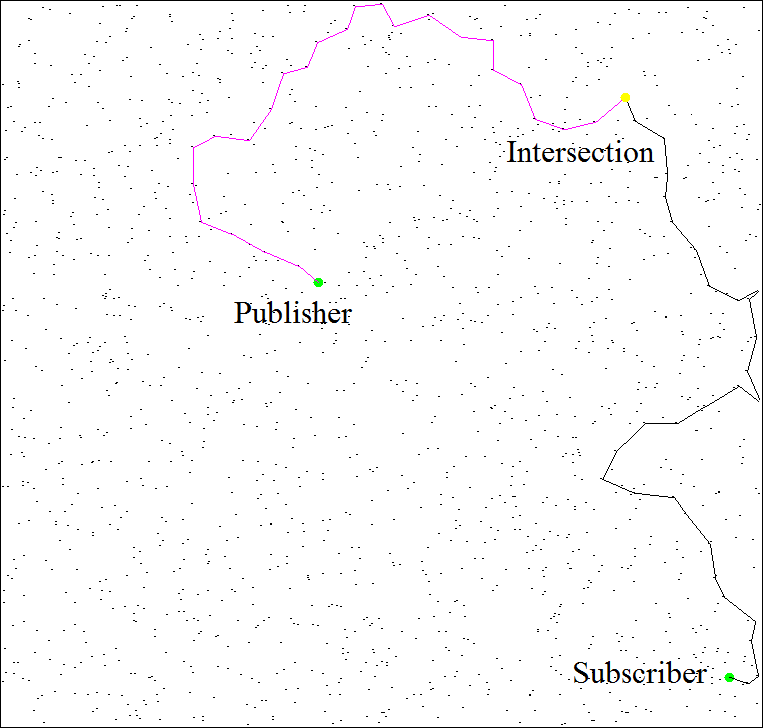}
\caption{Directional Random Walks intersecting using a Java simulator.}
\vspace{-1em}
\label{sim}
\end{figure}

A DRW is a probabilistic technique able to go forward into the network following a loop-free path. The principle assumed in this strategy is that two lines in a plane cross (see Figure~\ref{sim}). It is unclear how to construct a straight path of relaying nodes in ubiquitous unstructured networks without requiring global information and without making use of geo-coordinates. In this research, two different methods have been proposed in order to build DRWs \cite{ASCOMS13DRW}\cite{SENSORNETS14DRW}.


The matching of publishers and subscribers will be done using a special architecture of Bloom filters \cite{5751342} implemented at broker nodes. Bloom filters are probabilistic data structures that efficiently manage membership of a certain number of elements. The content related to membership is hashed using different hashing algorithms. Then, the positions of the Bloom structure corresponding to the hashes are set to one. The maximum number of elements to be inserted to the filter is fixed in order to maintain a certain probability of false positives. When searching for elements of a certain membership, the correspondent positions of the data structure are checked. The main advantage of Bloom filters is that they do not require much memory space and processing resources; so its use is very convenient in a sensing scenario in which devices have limited capabilities.

It is remarkable to mention that in our event-based system no advertisement table is required because filters just manage information about subscriptions.

\vspace{0.5em}
\section{Research efforts}
\label{sec:research}


In this section, we present the efforts already made in order to build the overlay layer proposed.

Based on \cite{ASCOMS13DRW}, a first method to build DRWs is proposed. It is based in the addition of different nodes to the DRW by pre-computing different weights at each node that take into account the two hops path. A weight is defined as follows:
\begin{equation}
\label{DRW Pierre}
n_{xz}^y = \vert \:  N(x)\cap N(z) \: \vert
\end{equation}
where $y$ is the last node added to the DRW; $x$ is the penultimate node added to the DRW; $z$ can be any node of the set $N(y)$ and $N(a)$ is the set of neighbor nodes of node $a$. Furthermore, in this method, a penalty is added to the weight when a node is added to the DRW.

Some properties about our heuristics were found using extensive simulations. The first property claims that DRWs decrease the time to intersection compared to pure random walks. The second property states that cooperation also decreases the time to intersection. Cooperation refers to synchronicity between publishers and subscribers. Finally, it is shown that DRWs are good at balancing the load of the network.

Based on \cite{SENSORNETS14DRW}, a second method to build DRWs is proposed. The main difference with the first design presented for DRWs is that nodes of the first and second neighborhoods of nodes added to the DRW are marked. In addition to this, the cost is not pre-computed, but it is computed when selecting a node as follows:
\begin{equation}
\label{DRW Cristina}
c(v) = \alpha \vert N(v) \cap N(DRW) \vert + \beta \vert N(v) \cap N^2(DRW) \vert
\end{equation}
where $\alpha$ and $\beta$ are parameters used as weights; $v$ can be any node of the set related to the neighborhood of the last node added to the DRW; $DRW$ is the set of nodes that are part of the DRW; $N(a)$ is the set of neighbor nodes of node $a$ and $N^2(DRW)$ is the set of neighbor nodes of $N(DRW)$.

In the first part of this research, the properties associated with a DRW were assessed. Implementations of DRWs of one branch or two branches were studied. The main results show that the use of one branch is as efficient as the use of two branches. Moreover, it is shown that the use of second neighborhoods to forward the DRW does not improve the Euclidean distance traversed in the network. It is also shown that shorter paths are obtained when using higher densities of nodes in the network. In the second part of this research, an information brokerage system was evaluated using a double ruling method. As in the first paper, it is shown that the algorithm is good at balancing the load using a few nodes of the network. In fact, we can state that the method proposed is as good as a traditional Rumor Routing algorithm \cite{Braginsky:2002:RRA:570738.570742} with an infinite memory.

\vspace{0.5em}
\section{Conclusion and Future work}
\label{sec:conclusion}

In this paper, a  novel network architecture for distributed event-based systems that use sensing devices has been proposed. Our first results, validated through extensive simulations, show that DRWs are suitable for the construction of an overlay layer that provides point-to-point communication and a distributed notification service. This is due, mainly to the good properties of DRWs to balance the load using a few nodes of the network for the establishment of paths. 

\begin{figure}[!h]
\centering
\includegraphics[width=2.5in]{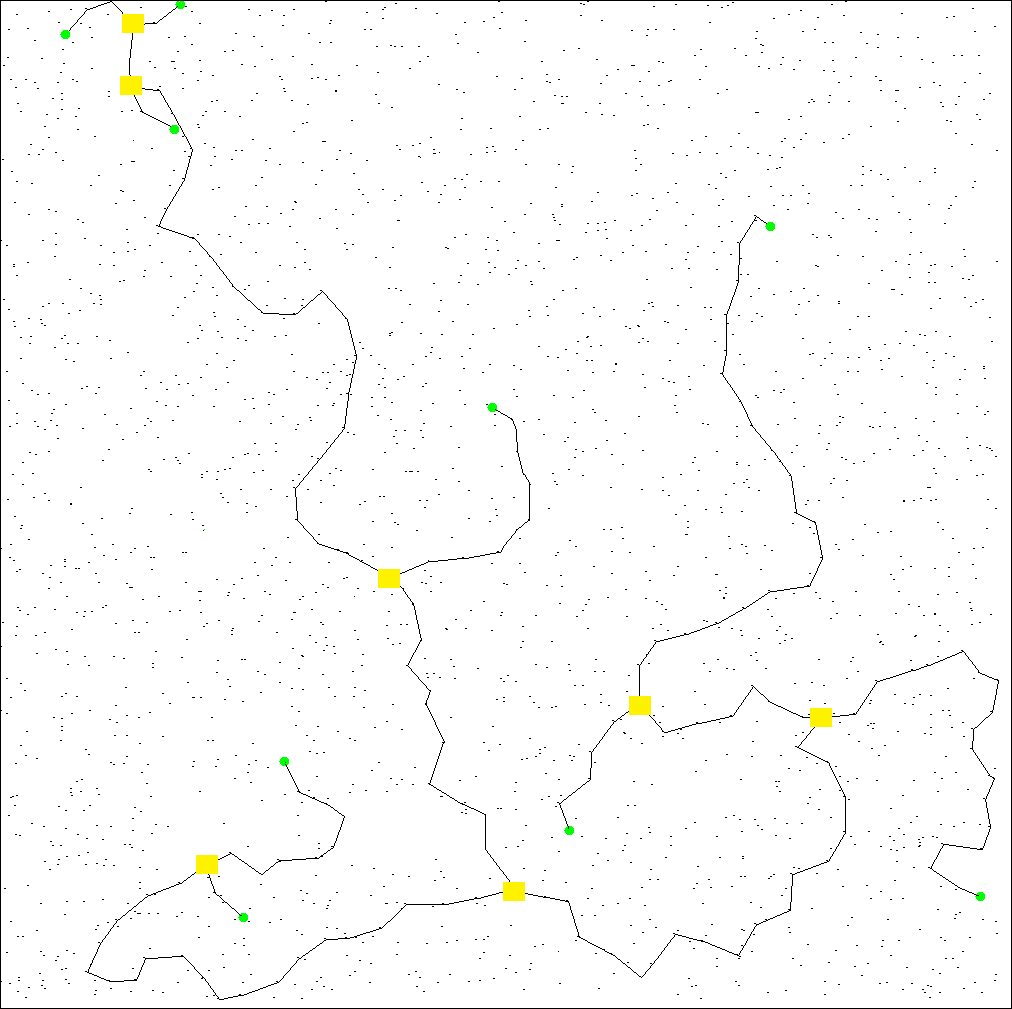}
\caption{Overlay layer using Directional Random Walks in a Java simulator.}
\vspace{-1.02em}
\label{overlaylayer}
\end{figure}

Currently, we are working on the first simulations of the overlay layer to study the impact of having different number of publishers and subscribers. Figure~\ref{overlaylayer} shows a simulation of the overlay layer in which yellow squares represent the distributed network of brokers while publishers and subscribers are represented using green circles.

At the same time, we are working on the design of Bloom filters for broker nodes that will be specifically fitted for sensing constrained devices. 

Finally, the deployment of a sensing testbed that will use wireless sensor nodes, will evaluate the suitability of the proposed solution under real conditions.

\vspace{0.27em}
\section*{Acknowledgment}

This work has been developed as part of the POPWiN project (Parallel Object Remote Programming for Heterogeneous Wireless Networks over IPv6) that is financially supported by the Hasler Foundation in its “SmartWorld - Information and Communication Technology for a Better World 2020” program.

\bibliographystyle{IEEEtran}
\bibliography{ref}

\begin{thebibliography}{10}
\providecommand{\url}[1]{#1}
\csname url@samestyle\endcsname
\providecommand{\newblock}{\relax}
\providecommand{\bibinfo}[2]{#2}
\providecommand{\BIBentrySTDinterwordspacing}{\spaceskip=0pt\relax}
\providecommand{\BIBentryALTinterwordstretchfactor}{4}
\providecommand{\BIBentryALTinterwordspacing}{\spaceskip=\fontdimen2\font plus
\BIBentryALTinterwordstretchfactor\fontdimen3\font minus
  \fontdimen4\font\relax}
\providecommand{\BIBforeignlanguage}[2]{{%
\expandafter\ifx\csname l@#1\endcsname\relax
\typeout{** WARNING: IEEEtran.bst: No hyphenation pattern has been}%
\typeout{** loaded for the language `#1'. Using the pattern for}%
\typeout{** the default language instead.}%
\else
\language=\csname l@#1\endcsname
\fi
#2}}
\providecommand{\BIBdecl}{\relax}
\BIBdecl
\renewcommand{\BIBentryALTinterwordstretchfactor}{4}

\bibitem{Cook:2012:RPC:2109687.2109848}
D.~J. ̃Cook and S.~K. ̃Das, ``Review: Pervasive computing at scale:
  Transforming the state of the art,'' Pervasive Mob. Comput., vol.~8, no.~1,
  Feb. 2012, pp. 22--35.

\bibitem{Muhl:2006:DES:1162246}
G. ̃M\"{u}hl, L. ̃Fiege, and P. ̃Pietzuch, Distributed Event-Based
  Systems.\hskip 1em plus 0.5em minus 0.4em\relax Secaucus, NJ, USA:
  Springer-Verlag New York, Inc., 2006.

\bibitem{Roy05securingmaodv:}
S. ̃Roy, V. ̃Addada, S. ̃Setia, and S. ̃Jajodia, ``Securing maodv: attacks
  and countermeasures,'' in Sensor and Ad Hoc Communications and Networks,
  2005. IEEE SECON 2005. 2005 Second Annual IEEE Communications Society
  Conference on, Sept 2005, pp. 521--532.

\bibitem{shi2011tinymq}
K. ̃Shi, Z. ̃Deng, and X. ̃Qin, ``Tinymq: A content-based publish/subscribe
  middleware for wireless sensor networks,'' in SENSORCOMM 2011, The Fifth
  International Conference on Sensor Technologies and Applications, 2011, pp.
  12--17.

\bibitem{souto2006mires}
E. ̃Souto et~al., ``Mires: a publish/subscribe middleware for sensor
  networks,'' Personal and Ubiquitous Computing, vol.~10, no.~1, 2006, pp.
  37--44.

\bibitem{Fang05glider:gradient}
Q. ̃Fang, J. ̃Gao, L.~J. ̃Guibas, V. ̃Silva, and L. ̃Zhang, ``Glider:
  Gradient landmark-based distributed routing for sensor networks,'' in in
  Proc. of the 24th Conference of the IEEE Communication Society (INFOCOM,
  2005, pp. 339--350.

\bibitem{Fang06landmark-basedinformation}
Q. ̃Fang, J. ̃Gao, and L.~J. ̃Guibas, ``Landmark-based information storage
  and retrieval in sensor networks,'' in In The 25th Conference of the IEEE
  Communication Society (INFOCOM’06, 2006, pp. 1--12.

\bibitem{Pastry}
A.~I.~T. ̃Rowstron and P. ̃Druschel, ``Pastry: Scalable, decentralized object
  location, and routing for large-scale peer-to-peer systems,'' in Proceedings
  of the IFIP/ACM International Conference on Distributed Systems Platforms
  Heidelberg, ser. Middleware '01.\hskip 1em plus 0.5em minus 0.4em\relax
  London, UK, UK: Springer-Verlag, 2001, pp. 329--350.

\bibitem{CAN}
S. ̃Ratnasamy, P. ̃Francis, M. ̃Handley, R. ̃Karp, and S. ̃Shenker, ``A
  scalable content-addressable network,'' in Proceedings of the 2001 conference
  on Applications, technologies, architectures, and protocols for computer
  communications, ser. SIGCOMM '01.\hskip 1em plus 0.5em minus 0.4em\relax New
  York, NY, USA: ACM, 2001, pp. 161--172.

\bibitem{Fersi:2013:DHT:2429525.2429572}
G. ̃Fersi, W. ̃Louati, and M. ̃Ben~Jemaa, ``Distributed hash table-based
  routing and data management in wireless sensor networks: a survey,'' Wirel.
  Netw., vol.~19, no.~2, Feb. 2013, pp. 219--236.

\bibitem{SENSORNETS14M3}
H. ̃Mahkonen, T. ̃Jokikyyny, J. ̃Jim\'enez, and S. ̃Kukli\'nski, ``M3:
  Machine-to-machine management framework,'' in Proc. 3rd Intl. Conference on
  Sensor Networks, SENSORNETS, 2014, pp. 139--144.

\bibitem{Lee:2002}
S.~J. ̃Lee, W. ̃Su, and M. ̃Gerla, ``On-demand multicast routing protocol in
  multihop wireless mobile networks,'' Mob. Netw. Appl., vol.~7, no.~6, Dec.
  2002, pp. 441--453.

\bibitem{Yoneki:2004}
E. ̃Yoneki and J. ̃Bacon, ``An adaptive approach to content-based
  subscription in mobile ad hoc networks,'' in Proceedings of the Second IEEE
  Annual Conference on Pervasive Computing and Communications Workshops, ser.
  PERCOMW '04.\hskip 1em plus 0.5em minus 0.4em\relax Washington, DC, USA: IEEE
  Computer Society, 2004, pp. 92--97.

\bibitem{Mottola:2008}
L. ̃Mottola, G. ̃Cugola, and G.~P. ̃Picco, ``A self-repairing tree topology
  enabling content-based routing in mobile ad hoc networks,'' IEEE Transactions
  on Mobile Computing, vol.~7, no.~8, Aug. 2008, pp. 946--960.

\bibitem{Voulgaris06}
S. ̃Voulgaris, E. ̃Rivière, A.-M. ̃Kermarrec, and M.~V. ̃Steen,
  ``Sub-2-sub: Self-organizing content-based publish subscribe for dynamic
  large scale collaborative networks,'' in In IPTPS’06: the fifth
  International Workshop on Peer-to-Peer Systems, 2006.

\bibitem{Tian:2005}
R. ̃Tian et~al., ``Hybrid overlay structure based on random walks,'' in
  Proceedings of the 4th international conference on Peer-to-Peer Systems, ser.
  IPTPS'05.\hskip 1em plus 0.5em minus 0.4em\relax Berlin, Heidelberg:
  Springer-Verlag, 2005, pp. 152--162.

\bibitem{Abdennadher:2013:APM:2508222.2508234}
F. ̃Abdennadher and M. ̃Ben~Jemaa, ``Accurate prediction of mobility into
  publish/subscribe,'' in Proceedings of the 11th ACM International Symposium
  on Mobility Management and Wireless Access, ser. MobiWac '13.\hskip 1em plus
  0.5em minus 0.4em\relax New York, NY, USA: ACM, 2013, pp. 101--106.

\bibitem{Haillot:2008}
J. ̃Haillot and F. ̃Guidec, ``Content-based communication in disconnected
  mobile ad hoc networks,'' in Proceedings of the 8th international conference
  on New technologies in distributed systems, ser. NOTERE '08.\hskip 1em plus
  0.5em minus 0.4em\relax New York, NY, USA: ACM, 2008, pp. 21:1--21:12.

\bibitem{1437119}
P. ̃Costa and G. ̃Picco, ``Semi-probabilistic content-based
  publish-subscribe,'' in Distributed Computing Systems, 2005. ICDCS 2005.
  Proceedings. 25th IEEE International Conference on, 2005, pp. 575--585.

\bibitem{Wong:2008}
B. ̃Wong and S. ̃Guha, ``Quasar: a probabilistic publish-subscribe system for
  social networks,'' in Proceedings of the 7th international conference on
  Peer-to-peer systems, ser. IPTPS'08.\hskip 1em plus 0.5em minus 0.4em\relax
  Berkeley, CA, USA: USENIX Association, 2008, pp. 2--2.

\bibitem{5751342}
S. ̃Tarkoma, C. ̃Rothenberg, and E. ̃Lagerspetz, ``Theory and practice of
  bloom filters for distributed systems,'' Communications Surveys Tutorials,
  IEEE, vol.~14, no.~1, First 2012, pp. 131--155.

\bibitem{Ramaswamy:2011}
L. ̃Ramaswamy and J. ̃Chen, ``The coquos approach to continuous queries in
  unstructured overlays,'' IEEE Trans. on Knowl. and Data Eng., vol.~23, no.~3,
  Mar. 2011, pp. 463--478.

\bibitem{ASCOMS13DRW}
P. ̃Leone and C. ̃Mu\~{n}oz, ``Content based routing with directional random
  walk for failure tolerance and detection in cooperative large scale wireless
  networks,'' in Proc. 2nd Intl. Workshop on Architecting Safety in
  Collaborative Mobile Systems, SAFECOMP, 2013, pp. 313--324.

\bibitem{SENSORNETS14DRW}
C. ̃Mu\~{n}oz and P. ̃Leone, ``Design of an unstructured and free
  geo-coordinates information brokerage system for sensor networks using
  directional random walks,'' in Proc. 3rd Intl. Conference on Sensor Networks,
  SENSORNETS, 2014, pp. 205--212.

\bibitem{Braginsky:2002:RRA:570738.570742}
D. ̃Braginsky and D. ̃Estrin, ``Rumor routing algorthim for sensor
  networks,'' in Proceedings of the 1st ACM international workshop on Wireless
  sensor networks and applications, ser. WSNA '02.\hskip 1em plus 0.5em minus
  0.4em\relax New York, NY, USA: ACM, 2002, pp. 22--31.

\end{thebibliography}

\end{document}